# Hot-carrier photocatalysts with energy-selective contacts based on quantum wells and dots

Shuanglong Han[a], Zhiqiang Fan[a], Ousi Pan[a], Xiaohang Chen[a], Zhimin Yang[b], Yanchao Zhang[c], Jincan Chen[a], Shanhe Su[a,1]

[a] *Department of Physics, Xiamen University, Xiamen 361005, People's Republic of China.*

[b] *School of Physics and Electronic Information, Yan'an University, Yan'an 716000, People's Republic of China*

[c] *School of Science, Guangxi University of Science and Technology, Liuzhou 545006, People's Republic of China*

**Abstract:**

In this paper, we simulate the function of hot-carrier photocatalysts (HCPCs) with quantum well and quantum dot energy-selective contacts (ESCs) in the water-splitting reaction. The transport equations for these ESCs are derived by using ballistic transport theory. The results indicate that thermalization loss from non-ideal ESCs is a primary factor diminishing the efficiency of HCPCs. The performance of HCPCs can be enhanced by optimizing the position of ESCs and the width of the extraction energy. Notably, HCPCs with quantum dot ESCs demonstrate superior performance compared to those with quantum well ESCs.

1 Email: sushanhe@xmu.edu.cn

## I. Introduction

Solar energy is a clean and limitless resource. However, sunlight is characterized by its variability and intermittency, making the direct utilization and storage of solar energy quite challenging.[1] Therefore, there is increasing interest in converting solar energy into other forms that can be stored and transported using various technologies.[2-4] In addition to solar cell technology, photocatalytic technology is also among the most promising options. This technology involves a semiconductor photocatalyst that absorbs photons with energy greater than its bandgap. This absorption generates photogenerated holes with oxidizing properties and photogenerated electrons with reducing properties, driving chemical reactions that are difficult to achieve under conventional conditions.[5] Photocatalytic technology can convert solar energy into chemical fuels such as hydrogen, which are easier to store and transport.[6-10] Additionally, it has the capability to degrade pollutants and facilitate environmental remediation.[11]

Electrons in the valence band of semiconductor catalysts absorb high-energy photons and transit to the conduction band, becoming high-energy electrons. However, the excess energy above the band gap in these high-energy electrons is quickly dissipated through thermalization.[12] Ultimately, these carriers migrate to the active surface sites, where they participate in specific chemical reactions. This process is accompanied by additional energy loss mechanisms, including both radiative and non-radiative recombination of electrons and holes. These energy loss mechanisms significantly limit the efficiency of conventional photocatalysts. Additionally, the absorption loss of sub-bandgap light further restricts their efficiency. For a photocatalyst to drive a chemical reaction, the chemical potential difference provided must be greater than the sum of the thermodynamic threshold voltage of the reaction and the electrocatalytic overpotentials of the two active sites.[13] This requirement generally makes conventional photocatalyst semiconductors with relatively large bandgaps. However, a larger bandgap also results in the increase of absorption losses. For example, $TiO_2$ photocatalyst can only absorb about $5\%$ of the solar spectrum in the ultraviolet range and are unable to utilize the much larger portion of the spectrum.[13] This low efficiency is a significant barrier to their practical applications.

Various strategies have been employed to enhance the efficiency of photocatalysts, including doping,[14,15] introducing co-catalysts,[16,17] and constructing heterojunction composite photocatalysts.[18-22] However, these methods often involve complex processes and high costs, and they are generally ineffective at reducing thermalization losses. Furthermore, the introduction of heterojunctions into photocatalysts can sometimes diminish their redox capabilities.[3] Yasuhiko et al. applied hot-carrier extraction to particulate photocatalysts to reduce thermalization losses,[23] and explored the effects of impact ionization and Auger recombination on hot-carrier photocatalysts (HCPCs).[24] Based on the analysis of HCPCs, we find that HCPCs not only reduce thermalization losses but also can utilize narrow-bandgap semiconductors to provide substantial redox capabilities. The narrower bandgap results in lower absorption losses. Theoretically, these catalysts can significantly enhance solar energy utilization, demonstrating tremendous potential.

In the field of solar cells, hot-carrier solar cells (HCSCs) are the counterpart to HCPCs. Research on HCSCs has been extensive,[25-27] with single-junction cells theoretically achieving efficiencies of up to $66\%$ or even higher.[28] For HCSCs and HCPCs to be realized, the carrier thermalization process must first be effectively slowed down from femtosecond time scales to nanosecond time scales, and second, high-energy carriers must be extracted efficiently. The mechanisms influencing the thermalization rate, including the phonon bottleneck effect (PBE), Auger recombination, and Auger relaxation, have been extensively studied across various types and structures of materials.[29,30] Li et al. discovered that carrier cooling lifetimes in halide perovskites can extend up to 1 nanosecond.[25] Efficient extraction of high-energy carriers requires the use of energy-selective contacts (ESCs). In semiconductor nanostructures, there are two common types of ESCs: quantum well ESCs and quantum dot ESCs. Both can be implemented using double-barrier resonant tunneling structures.[27,31] The quantum well ESC filters carriers based on momentum magnitude in a single direction,[32] while the quantum dot ESC filters carriers based on the total momentum magnitude.[33,34] The impact of the specific parameters of these two types of ESCs on HCPCs has not yet been explored, and the differences between the two types of HCPCs remain unclear.

In this paper, we optimize and analyze HCPCs with quantum well and quantum dot ESCs based on the water-splitting reaction. The differences between the two types of HCPCs are elucidated, and the proposed models demonstrate high energy conversion efficiency even when the band gap of the light-absorbing layer is less than $qV_0$. The rest of the paper is organized as follows: Section II introduces the proposed HCPC model. Section III presents the derivations for electron flow densities and energy fluxes extracted from the two types of ESCs. Section IV establishes the equilibrium equations and provides the efficiency expressions. Section V analyzes and discusses the simulation results. Finally, the paper concludes with a brief summary.

## Ⅱ. Modeling of hot-carrier photocatalysts

Fig. 1 illustrates the energy and structural diagrams of a HCPC model. The process of photogeneration leading to carrier extraction is similar to that of the HCSC. The absorber, serving as the core of the HCPC, is regarded as an ideal photovoltaic material.[23] Upon photoexcitation, the generated carriers collide with one another and rapidly undergo thermal relaxation, reaching an effective equilibrium state. This process takes place prior to thermalization with the lattice. In Fig. 1(a), $E_g$ represents the band gap of the light absorber, $T_H$ denotes the temperature of photogenerated carriers, $T_C$ indicates the temperature of the lattice, and $\Delta\mu = E_{fe} - E_{fh}$ signifies the difference between the chemical potentials $E_{fe}$ of electrons and $E_{fh}$ of holes in the absorber under illumination. The energy level at the center of the band gap of the absober is considered to be zero, serving as a reference point for measuring all other energy levels in the system.[12,35-37] While this is not a universal case, it provides a good approximation for capturing the physical properties of the problem. It is assumed that electrons and holes are symmetrically distributed according to the carrier temperature $T_H$. The chemical potentials for electrons and holes are $E_{fe} = -E_{fh} = \Delta\mu / 2$.

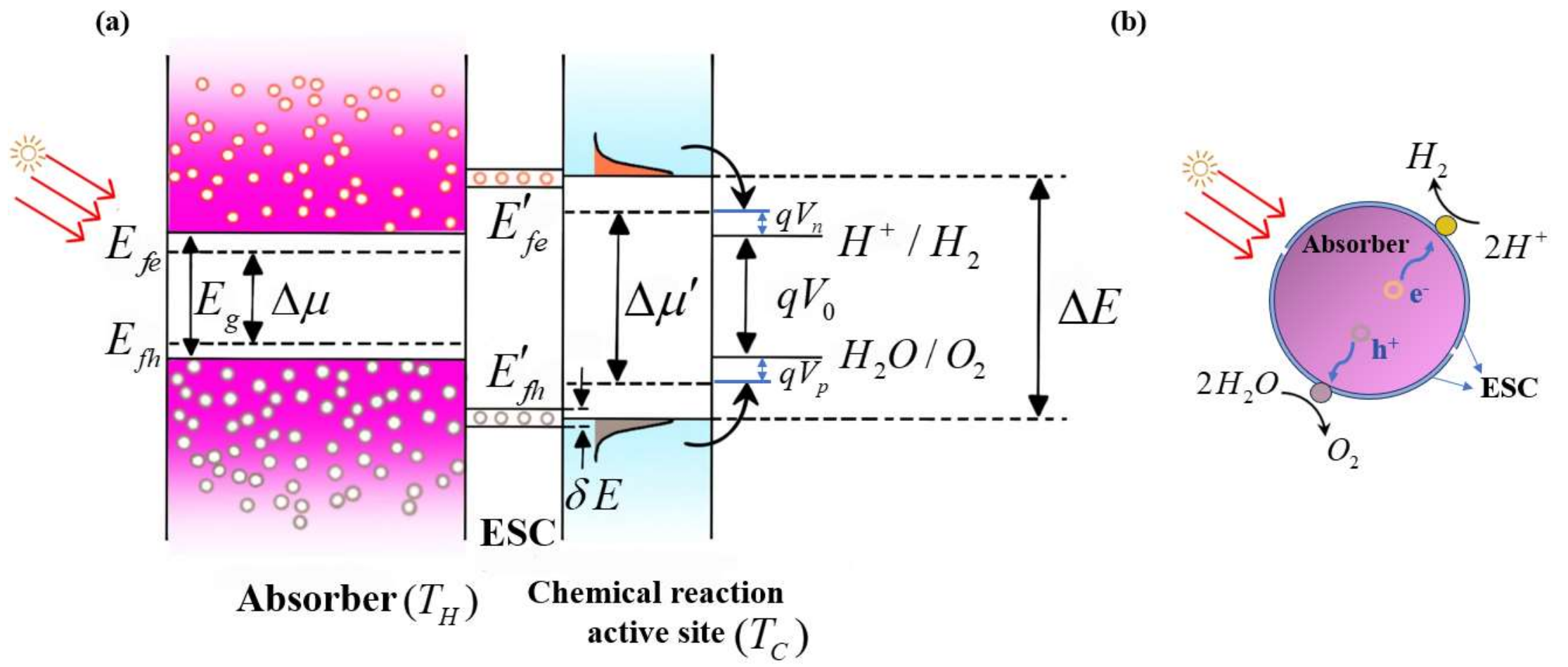

Fig. 1. (a) Energy diagram and (b) structural representation illustrating the mechanisms of a hot-carrier photocatalyst.

The hot-carrier absorber is encased by the shell, which functions as an ESC layer, as illustrated in Figs. 1(a) and 1(b). The photogenerated electrons and holes in the absorber are extracted through the ESC layer to the active sites from opposite faces. The distribution of particles in the chemical active site is characterized by temperature $T_C$. The ESC layer comprises multilayer heterostructures,[31] enabling hot carriers to transport within a very narrow energy range. Two rectangular energy transmission windows on opposite faces are utilized for electrons at the center energy level $\Delta E / 2$ and holes at $-\Delta E / 2$. $\delta E$ denotes the energy width for carrier extraction. $\Delta\mu' = E'_{fe} - E'_{fh}$ indicates the difference between the chemical potentials $E'_{fe}$ of electrons and $E'_{fh}$ of holes at the active site. Because of symmetrical distribution, $E'_{fe} = -E'_{fh} = \Delta\mu' / 2$. $V_0$ signifies the thermodynamic threshold voltage for chemical reactions, with specific values $V_0 = 1.23V$ for water splitting. $V_n$ and $V_p$ are the electrocatalytic overpotentials of the two reactive sites, respectively, and are assumed to be $V_n = V_p = V_{OV}$. In the following section of the article, we analyze the performance of the two types of hybrid catalytic photoelectrochemical systems (HCPCs) in the context of the hydrolysis-hydrogen production reaction. The chemical reactions occurring at the active site are as follows:

$$2h^+ + H_2O \rightarrow 2H^+ + \frac{1}{2}O_2 \tag{1}$$

and

$$2e^- + 2H^+ \rightarrow H_2, \tag{2}$$

where $h^+$ and $e^-$ denote holes and electrons, respectively.

## III. The transport equation of energy selective contacts

Energy-selective contacts (ESCs) facilitate the transport of particles between the absorber and the active site. Since the motions of electrons and holes are assumed to be symmetrical and equivalent, we focus exclusively on electron transport and derive the relevant equations. In Fig. 1, the distribution of electrons in the absorber and the active site is described by the Fermi-Dirac distribution functions $f\left(E, E_{fe}, T_H\right)$ and $f\left(E, E'_{fe}, T_C\right)$, respectively. $f(E,\mu,T) = \left\{1+\exp\left[\left(E-\mu\right)/\left(k_B T\right)\right]\right\}^{-1}$ is the Fermi-Dirac distribution function of the electron reservoir with temperature $T$ and chemical potential $\mu$, where $k_B$ denotes the Boltzmann constant.

Fig. 2 illustrates the Fermi spheres of electron transport for two types of energy-selective contacts (ESCs), highlighting the differences in momentum space between the transport mechanisms of the two device types. The shaded region in the figure represents the range of electron momenta that can be transported. For quantum well ESCs, electron transport depends solely on the momentum component of the wave vector $\vec{k}_x$. In contrast, for quantum dot ESCs, electrons can be transported as long as their total momentum exceeds the threshold $k'$, where $k'$ is the magnitude of the wave vector $\vec{k}'$ in the momentum space.[38]

By assuming ballistic electron transport, the electron flow density extracted from an electronic reservoir through a quantum well ESC is described by the Landauer equation [39]

$$\dot{N}^{QW} = 2\int_{-\infty}^{\infty}\int_{-\infty}^{\infty}\int_{0}^{\infty} f\left(E\left(\vec{k}\right),\mu,T\right)v\left(k_x\right)\xi\left(k_x\right)\frac{dk_x}{2\pi}\frac{dk_y}{2\pi}\frac{dk_z}{2\pi} \tag{3}$$

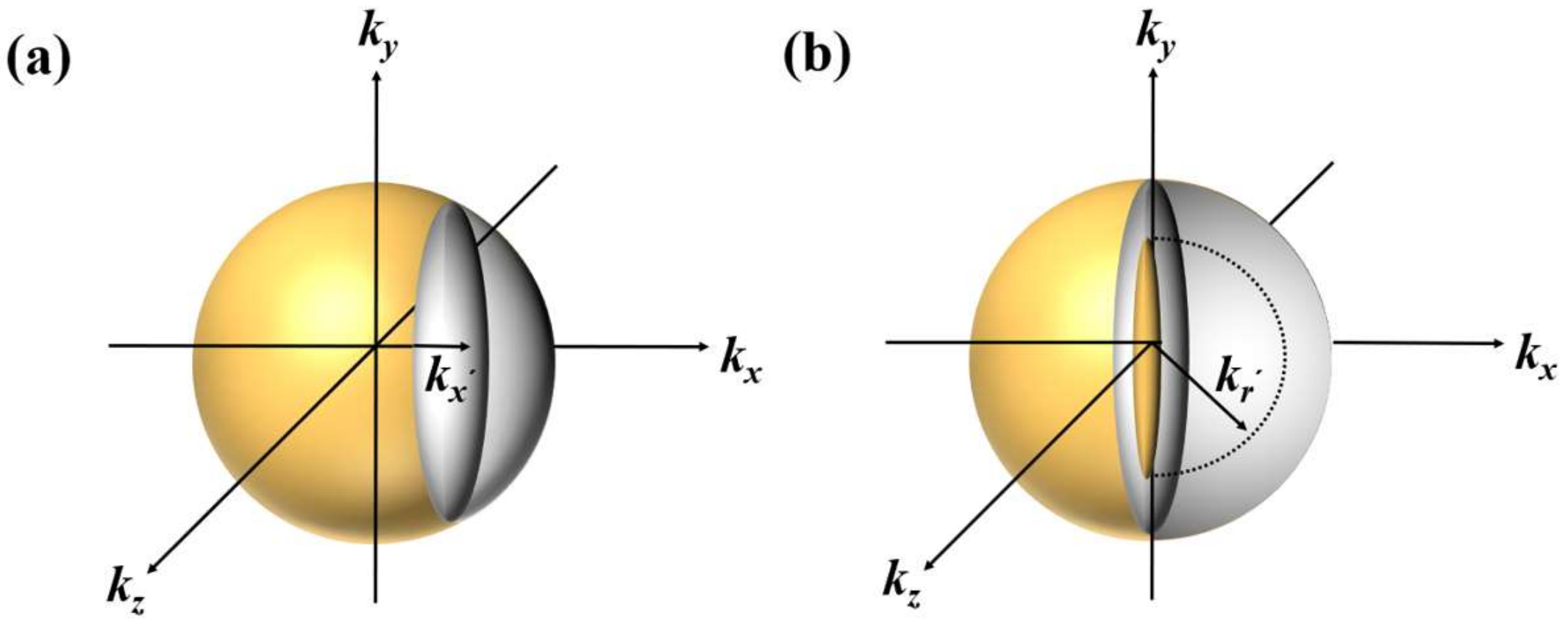

Fig. 2. Fermi spheres illustrating electrons transmitted through (a) quantum well energy-selective contacts and (b) quantum dot energy-selective contacts.

where $k_x$, $k_y$, and $k_z$ denote the magnitude of wave vectors in $x$, $y$, and $z$ directions, respectively, the factor 2 accounts for the spin degeneracy of electrons. $E\left(\vec{k}\right) = \hbar^2 k_x^{\ 2}/2m^* + \hbar^2 k_y^{\ 2}/2m^* + \hbar^2 k_z^{\ 2}/2m^*$ denotes the dispersion relation, where $\hbar = h/2\pi$ with $h$ being the Planck constant, $m^*$ denotes the effective mass of electrons. $v\left(k_x\right) = \hbar k_x / m^*$ is the velocity of the electrons in the x-direction, and $\xi\left(k_x\right)$ denotes the probability of the electrons being transmitted.

Through analytical calculations, Eq. (3) can be simplified to

$$\dot{N}^{QW} = \frac{1}{2\pi\hbar}\int_{0}^{\infty} n^{QW}\left(\mu,T\right)\xi\left(E_x\right)dE_x, \tag{4}$$

where

$$n^{QW}\left(\mu,T\right) = \frac{m^* k_B T}{\pi\hbar^2}\log\left[1+\exp\left(-\frac{E_x-\mu}{k_B T}\right)\right]. \tag{5}$$

In a HCPC system featuring a quantum well ESC, the net electron flow density extracted from the absorber to the quantum well ESC is determined by the difference between two components: i.e., the difference between the electron flow density moving from the absorber to the active site and the electron flow density moving from the active site back to the absorber, which is given by

$$\dot{N}_{net}^{QW} = \frac{1}{2\pi\hbar}\int_0^{\infty}\left[n^{QW}\left(E_{fe},T_H\right) - n^{QW}\left(E'_{fe},T_C\right)\right]\xi\left(E_x\right)dE_x . \tag{6}$$

According to the first law of thermodynamics, each electron that leaves or enters the reservoir carries energy equivalent to its total energy minus the electrochemical potential energy of the reservoir [40]. Under the Boltzmann approximation, for a quantum well ESC, the total energy per electron is the sum of the energy in the x-direction and the contributions from the other two degrees of freedom. Each of these additional degrees of freedom contributes $k_B T/2$ to the total energy. Thus, the energy transferred from the absorber to the active site per electron is $E_x + k_B T_H - E_{fe}$, while the energy transferred from the active site back to the absorber per electron is $E_x + k_B T_C - E'_{fe}$. Consequently, by combining with Eq. (6), the net energy flux extracted from the absorber by the quantum well ESC is given by

$$\dot{E}_{net}^{QW} = \frac{1}{2\pi\hbar}\int_0^{\infty}\left[\left(E_x + k_B T_H - E_{fe}\right)n^{QW}\left(E_{fe},T_H\right) - \left(E_x + k_B T_C - E'_{fe}\right)n^{QW}\left(E'_{fe},T_C\right)\right]\xi\left(E_x\right)dE_x . \tag{7}$$

For a quantum dot ESC, the electron flow density extracted from an electron reservoir can also be described by using the Landauer equation

$$\dot{N}^{QD} = 2\int_{-\infty}^{\infty}\int_{-\infty}^{\infty}\int_{0}^{\infty} f\left(E\left(\vec{k}\right),\mu,T\right)v\left(k_x\right)\xi\left(\vec{k}\right)\frac{dk_x}{2\pi}\frac{dk_y}{2\pi}\frac{dk_z}{2\pi} . \tag{8}$$

The difference between Eq. (3) and Eq. (8) lies in the dependence of the transfer function. In Eq. (3), $\xi\left(k_x\right)$ depends on $k_x$, while $\xi\left(\vec{k}\right)$ in Eq. (8) depends on the total momentum. By transforming from the momentum space to the energy space, Eq. (8) can be simplified as

$$\dot{N}^{QD} = \frac{m^*}{2\pi^2\hbar^3}\int_0^\infty Ef\left(E,\mu,T\right)\xi\left(E\right)dE\,. \tag{9}$$

Therefore, the net electron flow density extracted from the absorber by the quantum dot ESC is

$$\dot{N}_{net}^{QD} = \frac{m^*}{2\pi^2\hbar^3}\int_0^\infty E\left[f\left(E,E_{fe},T_H\right)-f\left(E,E_{fe}',T_C\right)\right]\xi\left(E\right)dE\,. \tag{10}$$

For a quantum dot ESC, the energy carried away or brought in by each electron leaving or entering the absorbing layer is $E-E_{fe}$. Thus, the net energy flux extracted from the absorber by the quantum dot ESC is

$$\dot{E}_{net}^{QD} = \frac{m^*}{2\pi^2\hbar^3}\int_0^\infty E\left(E-E_{fe}\right)\left[f\left(E,E_{fe},T_H\right)-f\left(E,E_{fe}',T_C\right)\right]\xi\left(E\right)dE\,. \tag{11}$$

The expressions for the particle flow density and net energy density at the hole contact are identical to those at the electron contact for both quantum well and dot ESCs. Specifically, $\dot{N}_{net,h}^{QW}=\dot{N}_{net,e}^{QW}$, $\dot{E}_{net,h}^{QW}=\dot{E}_{net,e}^{QW}$, $\dot{N}_{net,h}^{QD}=\dot{N}_{net,e}^{QD}$ and $\dot{E}_{net,h}^{QD}=\dot{E}_{net,e}^{QD}$ due to the assumption that the structures of the conduction and valence bands are approximately symmetric.

## IV. Conservation equations for particles and energy

Since the present article focuses on the effect of ESCs on the HCPCs, we assume that the absorber is ideal and carrier thermalization occurs slowly enough to permit carrier extraction before significant thermalization. Additionally, the radiative recombination of carriers is only considered.

For HCPCs, the electron flow density extracted through the energy-selective channel represents the carrier supply rate to the active site ($\dot{N}_{\sup}^{QW(QD)}$). This rate is equal to the difference between the carrier generation rate and the radiative recombination rate, i.e.,

$$\dot{N}_{\sup}^{QW(QD)} = \dot{N}_{net,e}^{QW(QD)} = \dot{N}_{net,h}^{QW(QD)} = \dot{G}^{QW(QD)} - \dot{R}^{QW(QD)}. \tag{12}$$

In this balance equation,

$$\dot{G}^{QW(QD)} = \int_{E_g}^{\infty} \frac{2\Omega}{h^3 c^2} \frac{E^2}{e^{E/(k_B T_S)} - 1} dE \tag{13}$$

denotes the carrier generation rate, which follows the generalized Planck law. Here, $c$ represents the speed of light and $T_S = 5760K$ is the temperature of the radiation source. $\Omega$ is the solid angle of solar radiation, which depends on the concentration ratio. A solar concentration ratio of 100 suns ($\Omega = 6.8 \times 10^{-3}$) is used in the following computation.

In addition,

$$\dot{R}^{QW(QD)} = \int_{E_g}^{\infty} \frac{2\pi}{h^3 c^2} \frac{E^2}{e^{(E-\Delta\mu)/(k_B T_H)} - 1} dE \tag{14}$$

describes the radiative recombination rate of carriers above the band gap, transitioning from the absorber to the environment in a planar geometry [39].

According to the principle of conservation of energy, the energy flux of solar radiation absorbed by the absorber $\dot{E}_G^{QW(QD)}$ equals the sum of the energy flux extracted from the absorber by the ESCs $\dot{E}_{net,e}^{QW(QD)} + \dot{E}_{net,h}^{QW(QD)}$ and the energy flux lost due to radiative recombination $\dot{E}_R^{QW(QD)}$, i.e.,

$$\dot{E}_G^{QW(QD)} = \dot{E}_{net,e}^{QW(QD)} + \dot{E}_{net,h}^{QW(QD)} + \dot{E}_R^{QW(QD)} \tag{15}$$

where the energy flux of solar radiation absorbed by the absorber $\dot{E}_G$ and the energy flux lost due to the radiative recombination $\dot{E}_R$ are, respectively, described as [39]

$$\dot{E}_G^{QW(QD)} = \int_{E_g}^{\infty} \frac{2\Omega}{h^3 c^2} \frac{E^3}{e^{E/(k_B T_S)} - 1} dE \tag{16}$$

and

$$\dot{E}_R^{QW(QD)} = \int_{E_g}^{\infty} \frac{2\pi}{h^3 c^2} \frac{E^3}{e^{(E-\Delta\mu)/(k_B T_H)} - 1} dE . \tag{17}$$

The electrocatalytic behavior at the active site of a chemical reaction is described by the Butler-Volmer equation [41]

$$j = j_0\left[\exp\left(\frac{\alpha_a V_{OV} q}{k_B T_C}\right) - \exp\left(-\frac{\alpha_c V_{OV} q}{k_B T_C}\right)\right]. \tag{18}$$

where $j$ is the current density supplied to the active site ( $j = q\dot{N}_{\text{sup}}^{QW(QD)}$ with $q$ representing the elementary charge), $j_0$ is the exchange current density, $\alpha_a$ and $\alpha_c$ are the anodic and cathodic charge transfer coefficients, respectively, and $V_{OV}$ is the overpotential. In the HCPC model, $qV_{OV} = (\Delta\mu' - qV_0)/2$.

The efficiency of energy conversion from solar energy to hydrogen is expressed by the following equation

$$\eta^{QW(QD)} = \frac{q \times \dot{N}_{\text{sup}}^{QW(QD)} \times 1.23(V)}{\dot{E}_{in}}, \tag{19}$$

where $\dot{E}_{in}$ represents the total energy flux of solar radiation and is calculated by

$$\dot{E}_{in} = \int_0^\infty \frac{2\Omega}{h^3 c^2} \frac{E^3}{e^{E/(k_B T_S)} - 1} dE. \tag{20}$$

It is important to note that in subsequent simulations, when examining the relationship between the device's carrier supply rate $\dot{N}_{\text{sup}}^{QW(QD)}$ and the chemical potential difference $\Delta\mu'$ at the active site, $\Delta\mu'$ is given directly, and $\Delta\mu$ and $T_H$ are obtained by solving the coupled equations consisting of Eqs. (12) and (15). In all other simulations, $\Delta\mu, T_H$, and $\Delta\mu'$ are obtained by solving the coupled equations from Eqs. (12), (15), and (18).

## V. Results and discussion

In the numerical calculations, the simulations were carried out using the following parameters. The effective mass of the electron was set to $m^* = 0.3m_e$, where $m_e$ is the mass of a free electron in the vacuum. The temperature of the active site was maintained at $300K$, and the electron transport probabilities for the quantum well and quantum dot ESCs were specified as follows:

$$\xi(E_x)=\begin{cases}1 & (\Delta E/2-\delta E/2\le E_x\le \Delta E/2+\delta E/2)\\ 0 & (otherwise)\end{cases} \tag{21}$$

and

$$\xi(E)=\begin{cases}1 & (\Delta E/2-\delta E/2\le E\le \Delta E/2+\delta E/2)\\ 0 & (otherwise)\end{cases}. \tag{22}$$

Other parameters will be provided in detail later.

Fig. 3 (a) [(b)] illustrates how the carrier supply rate $\dot{N}_{\sup}^{QW}$ ($\dot{N}_{\sup}^{QD}$) extracted through the quantum well (dot) ESC varies with the chemical potential difference $\Delta\mu'$ at the active site for different extraction energy $\Delta E$, where $E_g = 1eV$ and $\delta E = 0.01eV$. It can be observed that for a given $\Delta E$, when $\Delta\mu'$ is small, $\dot{N}_{\sup}^{QW}$ ($\dot{N}_{\sup}^{QD}$) remains high and relatively constant, which can be regarded as the saturation supply rate. However, as $\Delta\mu'$ increases beyond a certain point, $\dot{N}_{\sup}^{QW}$ ($\dot{N}_{\sup}^{QD}$) decreases sharply until it ultimately reaches zero. This is because for a given $\Delta E$, when $\Delta\mu'$ is small, both the chemical potential $\Delta\mu$ and the carrier temperature $T_H$ in the absorber are low. This results in a small or even negligible radiative recombination rate and allows that most carriers can be extracted by ESCs. At this stage, the primary factor limiting the device's energy conversion efficiency is of thermalization within the non-ideal ESCs. As $\Delta\mu'$ increases, the radiative recombination losses increase, leading to a rapid decrease of the carrier supply rate.

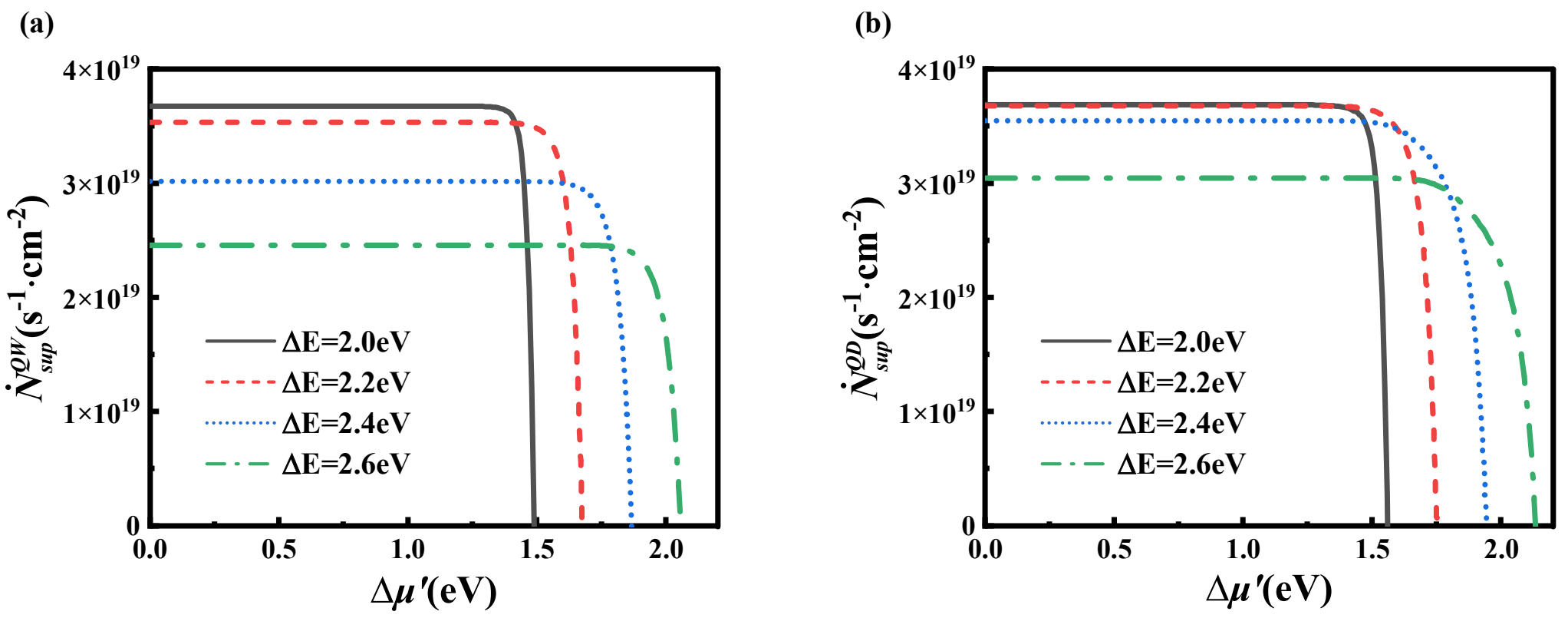

Fig. 3. The carrier supply rate $\dot{N}_{\sup}^{QW}$ ($\dot{N}_{\sup}^{QD}$) varying with the chemical potential difference $\Delta\mu'$ at the active site under different extraction energy $\Delta E$ for (a) HCPCs with quantum well ESCs and (b) HCPCs with quantum dot ESCs.

It is also observed in Fig. 3 that as the extraction energy $\Delta E$ increases, the range of $\Delta\mu'$ that corresponds to the carrier supply rate $\dot{N}_{\sup}^{QW} > 0$ ($\dot{N}_{\sup}^{QD} > 0$) also broadens. The hot-carrier photocatalyst can then drive chemical reactions that require a larger thermodynamic threshold voltage. Therefore, the extraction energy $\Delta E$ is an important parameter influencing the range of $\Delta\mu'$ that reflect the redox capability of the device. However, the increase of $\Delta E$ reduce the saturation supply rate of $\dot{N}_{\sup}^{QW}$ ($\dot{N}_{\sup}^{QD}$). Therefore, it is important to determine the appropriate value of $\Delta E$ according to the actual needs. By comparing Figs. 3(a) and (b), it is evident that under the same $\Delta E$, the HCPC with quantum dot ESCs outperforms the HCPC with quantum well ESCs in both the saturation supply rate and the range of $\Delta\mu'$ for positive carrier supply rate. This can be attributed to the fact that, as mentioned in Section III, quantum dot ESCs are capable of transporting a higher number of carriers.

The variation in solar-to-$H_2$ energy conversion efficiency $\eta^{QW}$ ($\eta^{QD}$) with extraction energy $\Delta E$ is shown in Fig. 4, where the parameters $\delta E = 0.01eV$, $\alpha_a = \alpha_c = 1.97$, and $j_0 = 0.147 mA \cdot cm^{-2}$.[41] For a given $E_g$, the efficiency generally increases and then decreases as $\Delta E$ increases. When $\Delta E$ is too small, the chemical potential difference $\Delta\mu'$ at the active site is not enough to drive the water decomposition reaction, and the energy conversion efficiency is zero. When $\Delta E$ is too large, less carriers can be extracted through ESCs, as few carriers are distributed at high energy level. The extraction energy $\Delta E$ of the ESC can be optimized to enhance the energy conversion efficiency. In addition, when $E_g$ is $1.1eV$ or $1.3eV$, there are certain regions where the variation of efficiency $\eta^{QW}$ ($\eta^{QD}$) with respect to $\Delta E$ is not very pronounced. A larger $E_g$ leads to a smaller chemical potential $\Delta\mu$ in the light absorber and a reduced value of radiative recombination rate $\dot{R}^{QW}$ ( $\dot{R}^{QD}$) according to Eq. (14). As a result, the carrier supply rate $\dot{N}_{\text{sup}}^{QW}$ ( $\dot{N}_{\text{sup}}^{QD}$) to the active site is approximately equal to the carrier generation rate $\dot{G}^{QW}$ ($\dot{G}^{QD}$)[Eq. (12)]. The variable trend in efficiency $\eta^{QW}$ ($\eta^{QD}$) and the carrier supply rate $\dot{N}_{\text{sup}}^{QW}$ ( $\dot{N}_{\text{sup}}^{QD}$) are consistent with each other [Eq. (19)].

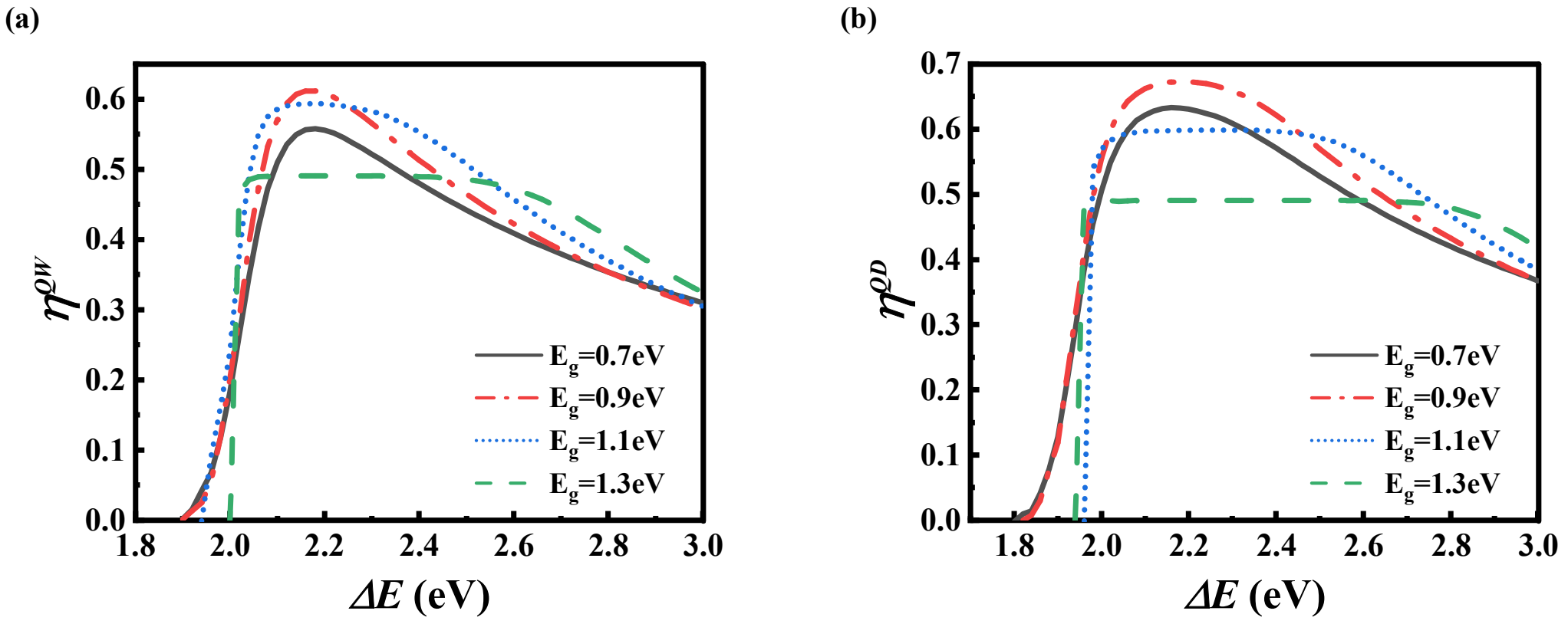

Fig. 4. The variation curves of the solar-to-H2 energy conversion efficiency $\eta^{QW}$ ($\eta^{QD}$) with the extraction energy $\Delta E$ under different energy band gap $E_g$ for (a) the HCPC with quantum well ESCs and (b) the HCPC with quantum dot ESCs.

Fig. 5 displays that the energy conversion efficiency $\eta^{QW}$ ($\eta^{QD}$) can also be improved by optimizing the energy band gap $E_g$. It can be explained by examining the curves of the carrier generation rate $\dot{G}^{QW}$ ($\dot{G}^{QD}$), radiative recombination rate $\dot{R}^{QW}$ ($\dot{R}^{QW}$), and carrier extraction rate $\dot{N}_{\text{sup}}^{QW}$ ($\dot{N}_{\text{sup}}^{QD}$) as they vary with the energy band gap $E_g$, all of which are also plotted in Fig. 5. As shown in Fig. 5, both the carrier generation rate $\dot{G}^{QW}$ ($\dot{G}^{QD}$) and the radiative recombination rate $\dot{R}^{QW}$ ($\dot{R}^{QW}$) decrease as $E_g$ increases. The radiative recombination rate $\dot{R}^{QW}$ ($\dot{R}^{QW}$) initially declines faster than the carrier generation rate $\dot{G}^{QW}$ ($\dot{G}^{QD}$), then decreases more slowly and eventually stabilizes. According to the particle conservation equation in Eq. (12), the carrier supply rate to the active site $\dot{N}_{\text{sup}}^{QW}$ ($\dot{N}_{\text{sup}}^{QD}$) first increases and then decreases. Notably, both Figs. (4) and (5) demonstrate that HCPC with quantum dot ESCs outperforms HCPC with quantum well ESCs under the same conditions.

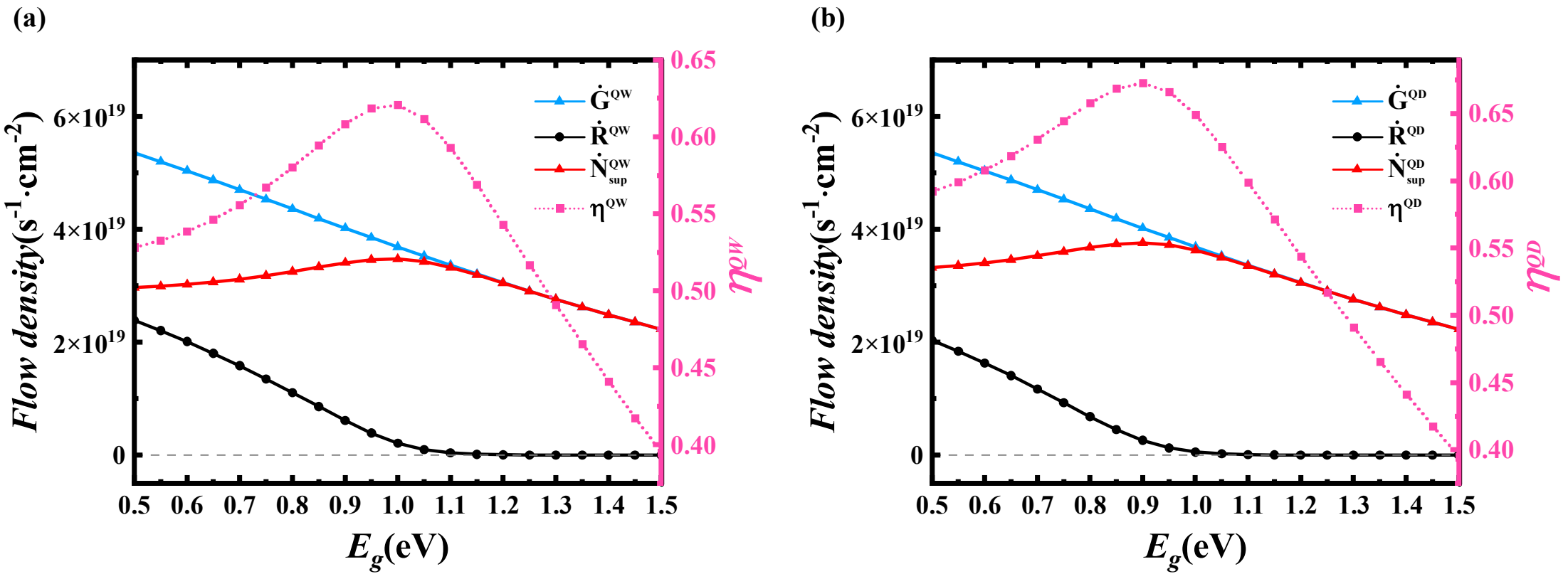

Fig. 5. The carrier generation rate $\dot{G}^{QW}$ ($\dot{G}^{QD}$), radiative recombination rate $\dot{R}^{QW}$ ($\dot{R}^{QD}$), carrier extraction rate $\dot{N}_{\text{sup}}^{QW}$ ($\dot{N}_{\text{sup}}^{QD}$) and efficiency $\eta^{QW}$ ($\eta^{QD}$) varying with the energy band gap $E_g$, where $\Delta E = 2.2eV$ and $\delta E = 0.01eV$ for (a) the HCPC with quantum well ESCs and (b) the HCPC with quantum dot ESCs. Note that the right vertical axis displays the values of efficiency, while the left vertical axis shows the values of the flow densities $\dot{G}^{QW}$ ($\dot{G}^{QD}$), $\dot{R}^{QW}$ ($\dot{R}^{QD}$), and $\dot{N}_{\text{sup}}^{QW}$ ($\dot{N}_{\text{sup}}^{QD}$).

Next, we will analyze why HCPCs with quantum dot ESCs demonstrate relatively superior performance. Fig. 6 illustrates how the energy flux density varies with $E_g$, highlighting the effects of the radiation recombination loss $\dot{E}_R^{QW}$ ( $\dot{E}_R^{QD}$ ) in the absorber and the thermalization $\dot{E}_{loss}^{QW}$ ( $\dot{E}_{loss}^{QD}$ ) in the ESCs, where $\Delta E = 2.2eV$ and $\delta E = 0.01eV$. The density of energy flow lost due to thermalization in the non-ideal is calculated by the following equation, i.e., $\dot{E}_{loss}^{QW(QD)} = \dot{E}_{net,e}^{QW(QD)} + \dot{E}_{net,h}^{QW(QD)} - \dot{N}_{\sup}^{QW(QD)} \times \Delta\mu'$. Fig. 6 illustrates that, for both types of HCPCs, the energy loss due to thermalization (dash curves) from non-ideal ESCs is significantly greater than the loss from radiative recombination (solid curves). This indicates that thermalization within the ESCs is one of the important factors affecting the performance of the device. The comparison reveals that quantum well ESCs have a higher thermalization energy loss rate $\dot{E}_{loss}^{QW}$ (black dash curve) compared to $\dot{E}_{loss}^{QD}$ of quantum dot ESCs (red dash curve) under the same parameter conditions. Moreover, HCPCs with quantum well ESCs lose more radiant energy $\dot{E}_R^{QW}$ (black solid curve) in the absorber layer compared to $\dot{E}_R^{QD}$ of quantum dot ESCs (red solid curve). Therefore, the performance of HCPCs with quantum dot ESCs is expected to surpass that of HCPCs with quantum well ESCs.

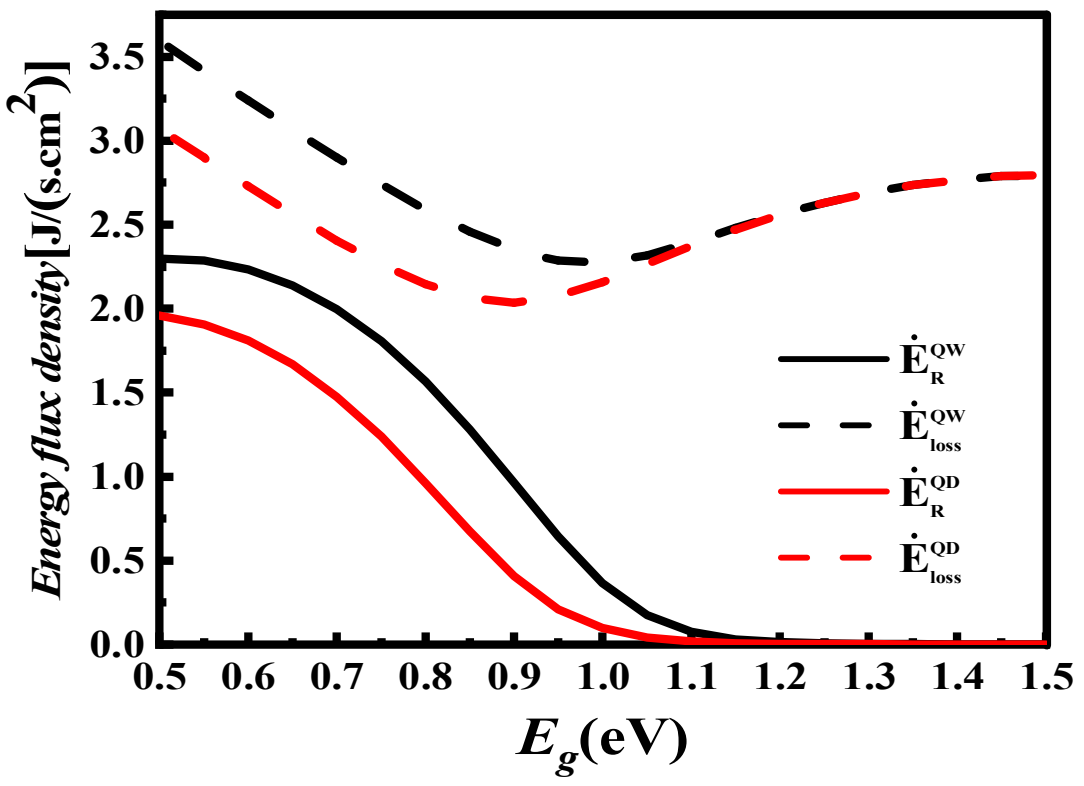

Fig. 6. Energy flux densities associated with the radiative recombination losses $\dot{E}_R^{QW}$ ( $\dot{E}_R^{QD}$ ) and the thermalization $\dot{E}_{loss}^{QW}$ ( $\dot{E}_{loss}^{QD}$ ) in the non-ideal ESCs as a function of the energy band gap $E_g$.

Figs. 7(a) and (b) display three-dimensional projections illustrating the solar-to-$H_2$ energy conversion efficiency $\eta^{QW}$ of HCPCs with quantum well and $\eta^{QD}$ of HCPCs with quantum dot ESCs. These projections are depicted as functions of $\delta E$ and $\Delta E$, where $E_g = 1eV$. From Figs. 7(a) and (b), it can be seen that $\eta^{QW}$ and $\eta^{QD}$ are not a monotonic function of $\delta E$ for a given $\Delta E$. When $\delta E$ is very small, the carrier extraction process through ESCs is hindered, resulting in high electron-hole recombination rates in the absorber. On the other hand, if $\delta E$ is too large, the ability of ESCs to select carriers decreases. This results in a greater increase in entropy and higher thermal losses due to the carrier extraction process in the non-ideal ESCs, ultimately reducing efficiency. By optimizing $\delta E$ and $\Delta E$, HCPCs with quantum dot ESCs achieve a maximum solar-to-$H_2$ energy conversion efficiency of 64.93% at $\Delta E = 2.22eV$ and $\delta E = 0.0085eV$, while HCPCs with quantum well ESCs reach a maximum efficiency of 62.34% at $\Delta E = 2.2eV$ and $\delta E = 0.191eV$. In addition, the better performance of the HCPC with quantum dot ESCs is further confirmed by comparing Fig. 7(a) and (b).

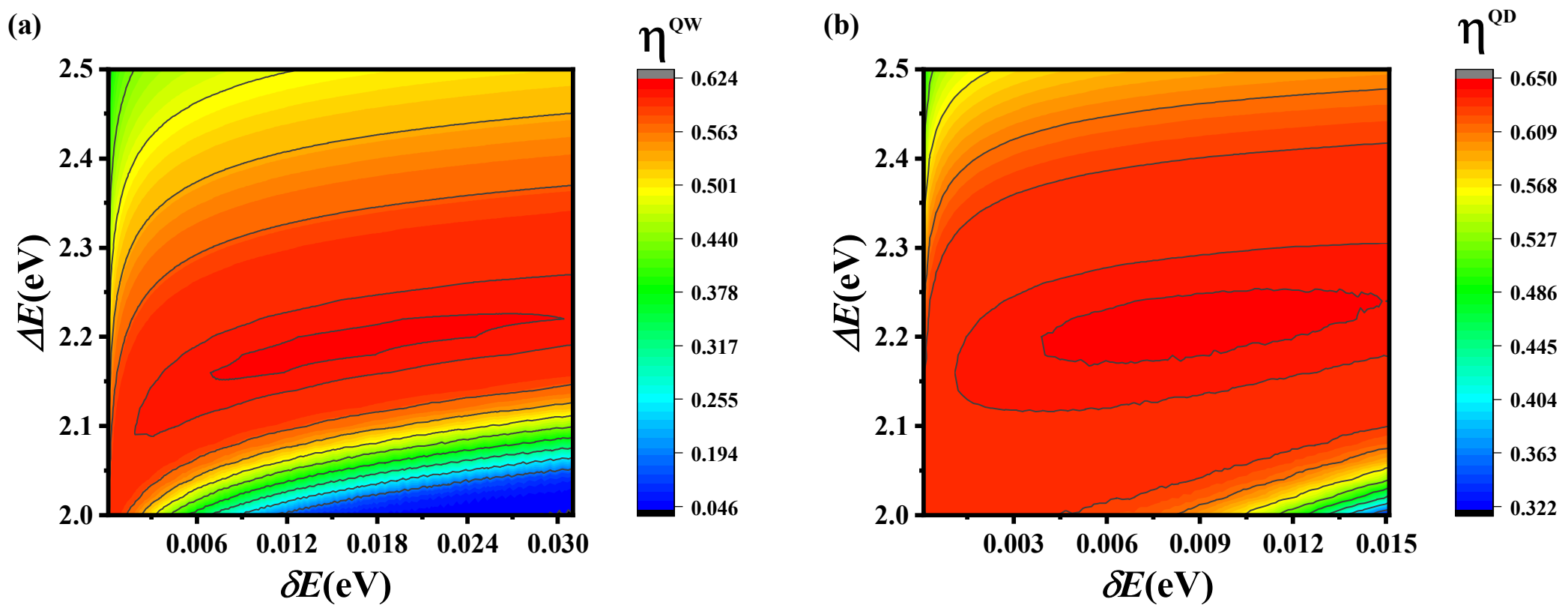

Fig. 7. The three dimensional graph of the solar-to-$H_2$ energy conversion efficiency $\eta^{QW}$ ($\eta^{QD}$) varying with the width of the ESC $\delta E$ and the extraction energy $\Delta E$ for (a) the HCPC with quantum well ESCs and (b) the HCPC with quantum dot ESCs.

## VI. Conclusions

We proposed two new types of HCPCs, derived the transport equations for both quantum well ESCs and quantum dot ESCs, and compared the performance of HCPCs equipped with these two types of ESCs in the context of water-splitting hydrogen production. It was found that, in addition to radiative recombination, energy loss from thermalization due to non-ideal ESCs is a primary factor limiting the energy conversion efficiency of HCPCs. By optimizing the parameters of ESCs, it was determined that HCPCs with quantum dot ESCs exhibit superior performance compared to those with quantum well ESCs. When the absorber has a narrower band gap ($E_g = 1eV$) than conventional photocatalysts, the maximum solar-to-$H_2$ energy conversion efficiencies achieved by HCPCs with quantum well ESCs and quantum dot ESCs are $62.34\%$ and $64.93\%$, respectively. Although the present paper is based on several simplified assumptions, it provides valuable insights into the mechanisms and performance of HCPCs with different types of ESCs.

**Acknowledgments**

This work has been supported by the National Natural Science Foundation of China (12364008, 12365006), Natural Science Foundation of Fujian Province (2023J01006), Fundamental Research Fund for the Central Universities (20720240145, 20720230012), and Natural Science Foundation of Guangxi Province (2022GXNSFBA035636).